\documentstyle[12pt]{article}
\begin{document}
\begin{flushright}
{BIHEP-TH-96-09}
\end{flushright}
\begin{center}
\section*{The perturbative pion-photon transition form factors 
with transverse momentum corrections} 

\vspace{1.5cm}

Fu-Guang Cao,
Tao Huang,
and Bo-Qiang Ma
\\
CCAST (World Laboratory), P.O.Box 8730, Beijing 100080, China and 
Institute of High Energy Physics, Academia Sinica, 
P.O.Box 918(4), Beijing 100039, China\footnote{Mailing address.
Email address: caofg@bepc3.ihep.ac.cn.}
\end{center}

(Received 31 January 1996)


\begin{abstract}
We perform a perturbative QCD analysis of the 
quark transverse momentum effect on the pion-photon transition
form factors $F_{\pi \gamma}$ and $F_{\pi \gamma^*}$
in the standard light-cone formalism, with two phenomenological
models of wavefunction as the input of the non-perturbative aspect of the 
pion.
We point out that the transverse momentum dependence 
in both the numerator 
and the denominator of the hard scattering amplitude 
is of the same importance and 
should be considered consistently.
It is shown that 
after taking into account the quark transverse
momentum corrections,
the results obtained from
different model wavefunctions
are consistent with the available
experimental data at finite $Q^2$.

\end{abstract}

\vskip 1cm
{PACS number(s): 12.38.Bx, 12.39.Ki, 13.40.Gp, 14.40.Aq
\newpage
\noindent
{\bf I. Introduction}

The pion-photon transition form factor $F_{\pi \gamma} (Q^2)$ is
a simple example for the perturbative analysis to exclusive
processes and was first analysed by Lepage and Brodsky
\cite{Lepage}. They predicted $F_{\pi \gamma} (Q^2)$ 
by neglecting $k_\bot$ relative to $q_\bot$,
\begin{eqnarray}
F_{\pi \gamma} (Q^2)=\frac{2}{\sqrt{3} Q^2} 
\int \frac{[{\rm d} x]}{x_1 x_2}\phi_\pi(x)
\left[ 1+{\rm O}\left(\alpha_s,\frac{m^2}{Q^2}\right) \right],
\label{fLepage}
\end{eqnarray}
and $Q^2 F_{\pi \gamma} (Q^2)$ would be essentially constant 
as $Q^2\rightarrow \infty$.
This approximation would be valid if the wavefunction 
is
peaked at low $k_\bot$ ( $k_\bot$ is the transverse momentum of quark)
so that $x_1 x_2 Q^2$ in the hard scattering
amplitude dominates the denominator. However, at the end-point region
$x_i \rightarrow 0,1$ and $Q^2 \sim$ a few GeV$^2$ the wavefunction
does not guarantee the $k_\bot$ negligible. One should take into account
$k_\bot$ corrections from both the hard scattering amplitude and
the wavefunction.

Recently, Refs.~\cite{Kroll,Ong} calculated 
the $\pi$-$\gamma$ transition form factor 
within the covariant hard scattering approach including transverse 
momentum effects and Sudakov corrections \cite{Sudakov} 
by neglecting the quark 
masses, the mass of the pion meson and the $k_\bot$-dependence in the
numerator of $T_H$. Their results show that Sudakov suppression
in the form factor $F_{\pi \gamma}(Q^2)$
is less important than in other exclusive channels and the 
Chernyak-Zhitnitsky(CZ) wavefunction should be discarded 
by fitting the
experimental data. However, as we know, the $k_\bot$-dependence
of the wavefunction in Ref.~\cite{Kroll} is the same in the
different models and it may be difficult to draw a conclusion
which excludes the CZ wavefunction.
We will re-examine this problem in the present paper.

The light-cone formalism provides a convenient framework for
the relativistic description of hadrons in terms of quark and gluon
degrees of freedom, and for the application of perturbative QCD (pQCD)
to exclusive processes \cite{BHL,Brod}. 
In this formalism, the hadronic
wavefunction which describes the hadronic composite 
state
at a particular
$\tau$ is expressed in terms of 
a series of light-cone wavefunctions in Fock-state basis, 
\begin{eqnarray}
| \pi \rangle=\sum | q \bar q \rangle \psi_{q \bar q}+
\sum | q \bar q g \rangle \psi_{q \bar q g}+\cdots,
\label{Fock}
\end{eqnarray}
and the temporal evolution of the state is generated by the light-cone
Hamiltonian $H_{LC}=P^-=P^0-P^3$.
Furthermore the vacuum 
state in the light-cone Fock basis is an
exact eigen-state of the full Hamiltonian $H_{LC}$. Thus all bare quanta 
in a hadronic Fock state are part of the hadron 
(This point is very different from that in the equal-$t$ 
perturbative theory
in which the quantization is performed at a given time $t$).
Light-cone pQCD is very convenient for light-cone dominated processes.
For the detail quantization rules we refer to literatures 
\cite{Lepage,BHL,Brod,lc}.
The more important point for practical
calculation is that the contributions coming from higher Fock states are 
suppressed by $1/Q^n$, therefore we can employ only the valence state to 
the leading order for large $Q^2$.
In this paper, 
we analyze the quark transverse momentum effects on the pion-photon
transition form factors $F_{\pi \gamma}$ and $F_{\pi \gamma^*}$
at finite $Q^2$
in the standard light-cone formalism, with two phenomenological
models of wavefunction as the input of the non-perturbative 
aspect of the pion.
We demonstrate that 
the pQCD predictions with the different models 
of wavefunction are consistent
with the available experimental data
by taking into account
the quark transverse momentum.

\vskip 0.5cm
\noindent
{\bf II. The pion-photon transition form factors $F_{\pi \gamma}$ and
$F_{\pi \gamma^*}$}

The $\pi$-$\gamma$ transition form factor $F_{\pi \gamma}$ is defined
from the $\pi^0 \gamma \gamma^\ast$ vertex 
in the amplitude of
$e \pi \rightarrow e \gamma$,
\begin{eqnarray}
\Gamma_\mu =- i e^2 F_{\pi \gamma} \epsilon_{\mu \nu \alpha \beta}
p^\mu_\pi \epsilon^\alpha q^\beta,
\end{eqnarray}
where $p_\pi$ and $q$ are the momenta of the incident pion and the
virtual photon respectively, and $\epsilon$ is the polarization
vector of the final (on-shell) photon. We adopt the standard 
momentum assignment at the
``infinite-momentum" frame \cite{Lepage}
\begin{eqnarray}
p_\pi&=&(p^+,p^-,p_\bot)=(1,0,0_\bot),\nonumber \\
q&=&(0,q_\bot^2,q_\bot),
\end{eqnarray}
where $p^+$ is arbitrary. For simplicity we choose $p^+=1$,
and we have $q^2=-q_\bot^2=-Q^2$. 
Then the $F_{\pi \gamma}$ is given by 
\begin{eqnarray}
F_{\pi \gamma}(Q^2)=\frac{\Gamma^+}{-i e (\epsilon_\bot \times q_\bot)},
\end{eqnarray}
where $\epsilon=(0,0,\epsilon_\bot)$, $\epsilon_\bot \cdot q_\bot=0$
is chosen and $\epsilon_\bot\times q_\bot=\epsilon_{\bot 1} q_{\bot 2} 
+\epsilon_{\bot 2} q_{\bot 1}$.
Since the contributions coming from higher Fock states
are suppressed,
we take into account only the conventional
lowest Fock state of pion meson,
\begin{eqnarray}
\psi_{\pi}= \frac{\delta^a_b}{\sqrt{n_c}}\frac{1}{\sqrt{2}}
\left[ \frac{u_\uparrow \bar u_\downarrow
-u_\downarrow \bar u_\uparrow}{\sqrt 2}
- \frac{d_\uparrow \bar d_\downarrow
-d_\downarrow \bar d_\uparrow}{\sqrt 2} \right]
\frac{\psi(x_i,k_\bot)}{\sqrt{x_1 x_2}}.
\label{piwf}
\end{eqnarray}
The leading-order contribution to $F_{\pi \gamma}$ is calculated
from Fig.~1 in light-cone pQCD \cite{Lepage},
\begin{eqnarray}
&&F_{\pi \gamma}(Q^2)=\frac{\sqrt{n_c}(e_u^2-e_d^2)}{i(\epsilon_\bot \times
q_\bot)}\int_0^1[{\rm d}x] \int_0^\infty \frac{{\rm d}^2 k_\bot}{16 \pi^3}
\psi(x_i,k_\bot) \nonumber \\
&& \left[\frac{\bar v_\downarrow (x_2,-k_\bot)} 
{\sqrt{x_2}}\rlap /\epsilon
\frac{u_\uparrow (x_1,k_\bot+q_\bot)}{\sqrt{x_1}} 
\frac{\bar u_\uparrow (x_1,k_\bot+q_\bot)}{\sqrt{x_1}} 
\gamma^+
\frac{u_\uparrow (x_1,k_\bot)}{\sqrt{x_1}} 
\frac{1}{D} + (1 \leftrightarrow 2) \right],
\label{spinor}
\end{eqnarray}
where $[{\rm d} x]= {\rm d}x_1 {\rm d}x_2 \delta(1-x_1-x_2),$
$e_{u,d}$ are the quark charges in unites of $e$,
and  $D$ is the ``energy-denominator",
\begin{eqnarray}
D=q_\bot^2-\frac{(k_\bot+q_\bot)^2+m^2}{x_1}
-\frac{k_\bot^2+m^2}{x_2}.
\end{eqnarray}
The quark masses relative to $Q^2$ 
can be neglected since they are the current quark masses
in pQCD calculation. Thus Eq.~(\ref{spinor}) becomes
\begin{eqnarray}
F_{\pi \gamma}(Q^2)=2 \sqrt{n_c}(e_u^2-e_d^2)\int_0^1 [{\rm d}x]
\int \frac{{\rm d}^2 k_\bot}{16 \pi^3} \psi(x_i,k_\bot)
\times T_H(x_1,x_2,k_\bot),
\label{f1Cao}
\end{eqnarray}
where 
\begin{eqnarray}
T_H(x_1,x_2,k_\bot)=
\frac{q_\bot \cdot (x_2 q_\bot + k_\bot)}
{q_\bot^2 (x_2 q_\bot + k_\bot)^2} +(1 \leftrightarrow 2).
\label{THCao}
\end{eqnarray}
The leading behavior of $T_H$ (at large $Q^2$) is obtained
by neglecting $k_\bot$ relative to $x_i q_\bot$ \cite{Brod},
\begin{eqnarray}
T_H^{LO}(x_1,x_2,k_\bot)=\frac{1}{x_1 x_2 Q^2}.
\end{eqnarray}
Thus Eq.~(\ref{f1Cao}) become Eq.~(\ref{fLepage})
to the leading order.
The higher twist corrections to $T_H^{LO}$ take the forms of
$\left(\frac{k_\bot}{x_i Q}\right)^n$.
Eq.~(\ref{THCao}) tells us that there are two factors to contribute
for the $k_\bot$-dependence. One is from the pQCD hard scattering
amplitude $T_H(x_i,Q,k_\bot)$, and another one is from the 
non-perturbative wavefunction $\psi(x_i,k_\bot)$.
Although one hopes that the end-point behavior of the wavefunction
can guarantee the reliability of neglecting these higher 
twist corrections and
can suppress the end-point singularity,
these corrections may substantially modify
the predictions for $F_{\pi \gamma}$ at the momentum transfer $Q$ of
a few GeV, especially for the wavefunction with a milder suppression
factor in the end-point region.
It should be
emphasized that the $k_\bot$-dependence in the numerator and the
denominator of $T_H$ is of the same importance. Thus one
can not simply ignore the $k_\bot$ term in the numerator of $T_H$ and
Eq.~(\ref{THCao}) gives the complete expression in the
leading order.

The $\pi$-$\gamma^\ast$ transition form factor $F_{\pi \gamma^\ast}$ 
is extracted from the $\pi^0 \gamma^\ast \gamma^\ast$ vertex 
in the two-photon physics. Once again, we employ the standard
momentum assignment at the
``infinite-momentum" frame 
\begin{eqnarray}
p_\pi&=&(p^+,p^-,p_\bot)=(1,0,0_\bot),\nonumber \\
 q&=&(0,q_\bot^2-Q'^2,q_\bot), \nonumber \\
q'&=&(1,q_\bot^2-Q'^2,q_\bot), 
\end{eqnarray}
where $q$ and $q'$ are the momenta of the two photons respectively,
and $q^2=-q_\bot^2=-Q^2$, $q'^2=-Q'^2$. 
$F_{\pi \gamma^\ast}$ may be calculated from Fig.~1 by substituting
a virtual photon $\gamma^\ast$ for the on-shell photon $\gamma$,
which gives
\begin{eqnarray}
F_{\pi \gamma}(Q^2,Q'^2)&=& 2 \sqrt{n_c}(e_u^2-e_d^2)\int_0^1 [{\rm
d}x]
\int \frac{{\rm d}^2 k_\bot}{16 \pi^3} \psi(x_i,k_\bot) \nonumber \\
& &\left[ \frac{q_\bot \cdot (x_2 q_\bot + k_\bot)}
 {q_\bot^2 \left[(x_2 q_\bot + k_\bot)^2 +x_1 x_2 q'^2_\bot \right]}
+(1 \leftrightarrow 2) \right].
\label{f2Cao}
\end{eqnarray}
The leading order behavior of $F_{\pi \gamma^\ast}$ can be obtained
from Eq.~(\ref{f2Cao}) by
neglecting $k_\bot$ relative to $x_i q_\bot$ \cite{Lepage},
\begin{eqnarray}
F_{\pi \gamma^\ast}(Q^2,Q'^2)=2 \sqrt{n_c}(e_u^2-e_d^2)
\int_0^1 [{\rm d} x]\phi_\pi(x) 
\left[ \frac{1}{x_2 Q^2+x_1 Q'^2} + (1 \leftrightarrow 2) \right].
\label{f2Brod}
\end{eqnarray}
Similar to the $F_{\pi \gamma}$, Eq.~(\ref{f2Cao}) may substantially
modifies the predictions obtained from Eq.~(\ref{f2Brod}).

\vskip 0.5cm
\noindent
{\bf III. Numerical calculations}

In order to see the transverse momentum corrections,
we employ two models of wavefunction: 
(a) the Brodsky-Huang-Lepage (BHL) wavefunction \cite{BHL}
\begin{eqnarray}
\psi^{BHL}(x,k_\bot)= A \rm {exp} \left[ -\frac{k_\bot^2+m^2}
{8 \beta^2 x (1-x)} \right ],
\label{model1}
\end{eqnarray}
where $A=32 \rm {GeV}^{-1}$, $\beta=0.385$ GeV
and $m=289$ MeV \cite{Mabq};
(b) the CZ-like wavefunction \cite{CZ} 
\begin{eqnarray}
\psi^{CZ}(x,k_\bot)= A (1-2 x)^2 
\rm {exp} \left[ -\frac{k_\bot^2+m^2}{8 \beta^2 x (1-x)} \right],
\label{model2}
\end{eqnarray}
where $A=136 \rm {GeV}^{-1}$, $\beta=0.455$ GeV and $m=342$ MeV \cite{Mabq}.
These models express that the Fock state  wavefunction      
$\psi(x_i,k_\bot)$ in the infinite 
momentum frame depends on the off-shell energy variable             
$\varepsilon=\sum\limits_i^n \left(\frac{k_{\bot i}^2+m_i^2}{x_i}\right)$, 
which 
was pointed out in Ref.~\cite{BHL}.

Substituting the models (\ref{model1}) and (\ref{model2})
into Eqs.~(\ref{f1Cao}), (\ref{THCao}) and (\ref{f2Cao}), one can get 
the transverse momentum corrections to the pion-photon 
transition form factor.
The results of $F_{\pi \gamma}$ calculated with    
$\psi^{BHL}$ and $\psi^{CZ}$ are plotted in Fig.~2. The 
dashed curves are calculated from the 
hard scattering amplitude 
$T_H^{LO}$ 
in the leading order without the transverse momentum corrections     
(see Eq.~(\ref{fLepage})), and the constant predictions with the 
different wavefunctions can not describe the experimental data 
at momentum transfer of a few GeV$^2$ 
explicitly.
The solid curves are obtained from the complete
expression of $T_H$ (see Eq. (\ref{THCao}))
with the transverse momentum corrections. 
As expected, the 
higher twist correction are suppressed by $1/Q^2$ and 
the prediction approaches
to a constant which depends on the wavefunction at large $Q$ region.
The perturbative predictions are smaller than the experimental data,
especially for $Q^2$ of $1 \sim 3$ GeV$^2$, which supports the
suggestion that the higher order effects should provide some
contributions at experimental accessible momentum transfer and
become more important with $Q^2$ decreasing.
Although the asymptotic behaviors of $F_{\pi \gamma}$ predicted
from the BHL model and CZ-like model of wavefunction are
quit different, their predictions at finite $Q^2$ obtained
with transverse momentum corrections are consistent with
the experimental data. The reason is as following: 
There are two factors to affect the 
prediction with the CZ-like wavefunction. First, the
CZ-like model emphasizes the end-point region in a strong way, 
which enhance its prediction of $F_{\pi \gamma}$.
Second, the transverse 
momentum corrections become more important in the 
end-point region, which 
make its prediction decrease. 
Combining these two factors, 
the CZ-like model gives a very similar prediction
as the BHL model in the finite momentum transfer region.
Thus, neither of the two models of wavefunction
can be excluded by the available data of this exclusive process.

The results of $F_{\pi \gamma^*}$ calculated with    
$\psi^{BHL}$ and $\psi^{CZ}$ are plotted in Fig.~3. 
Once again, the higher twist corrections are suppressed by
$1/Q^2$ and provide more contributions as $Q^2$
decreasing.
The predictions of the two models are not different dramatically,
no matter the transverse momentum corrections are taken into account
or not, since the energy scale $Q'$ 
coming from the other virtual photon makes the hard scattering amplitude 
is not as singular as that in the case of $F_{\pi \gamma}$.
It is also difficult to exclude one of the two models of wavefunction
basing on $F_{\pi \gamma^*}$.
At present, the lack of experiment data make the examination
of higher twist effects in $F_{\pi \gamma^*}$ more complex
than that in $F_{\pi \gamma}$. 
But the future high-luminosity $e^+ e^-$ colliders in the
``$\tau$-charm factory" or ``$B$ factory" will make this 
examination feasible.

\vskip 0.5cm
\noindent
{\bf III. Summary}

In summary,
we emphasize again that 
the light-cone perturbative QCD is a natural framework to calculate
the large-momentum-transfer exclusive processes.
It is reasonable to get the higher twist corrections by taking
into account the quark transverse momentum dependence.
As $Q^2 \rightarrow \infty$, these corrections become negligible.
After taking into account the transverse momentum dependence,
pQCD may give correct prediction for the pion-photon transition
form factor which is consistent with the experimental data.
The transverse-momentum-dependence in both the numerator and
the denominator of the hard scattering amplitude $T_H$ is of the
same importance and should be considered consistently.
Neither the BHL model nor the CZ-like model, the two typical
models of wavefunctions, can be excluded by the available data of
the 
pion-photon transition form factors.
The future ``$\tau$-charm factory" as well as ``$B$ factory"
will provide the opportunity to examine the higher twist effects
in the perturbative calculation of $F_{\pi \gamma^*}$ and
to test the validity of the perturbative analysis.

\vskip 0.5cm
\begin{center}
{\bf Acknowledgments}
\end{center}
We would like to thank S.J. Brodsky and H.N. Li for helpful discussions.


\newpage
\section*{Figure Captions}
\begin{description}
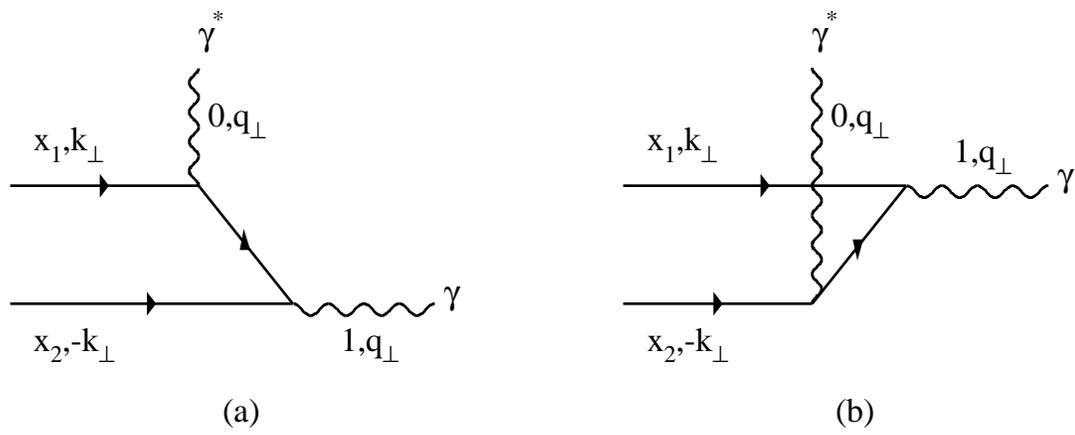
\item{Fig.~1} The lowest order diagrams contributing to $F_{\pi \gamma}$ 
in light-cone pQCD.
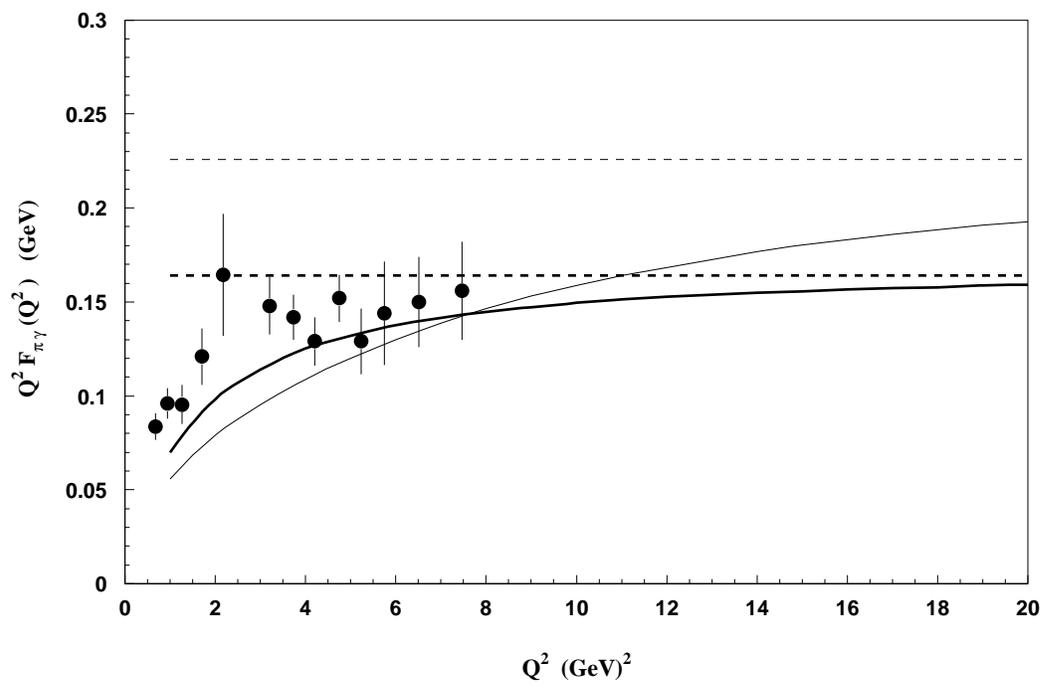
\item{Fig.~2} The $\pi$-$\gamma $ transition form factor. 
The solid curves are obtained by taking into account the
$k_\bot$-dependence, while the dashed curves are results without
$k_\bot$-dependence.
In both of the cases, the thick curves are calculated from the BHL wave
function and the thin curves are for the CZ-like wavefunction.
The data are taken from Refs.~\cite{data,cleo}.
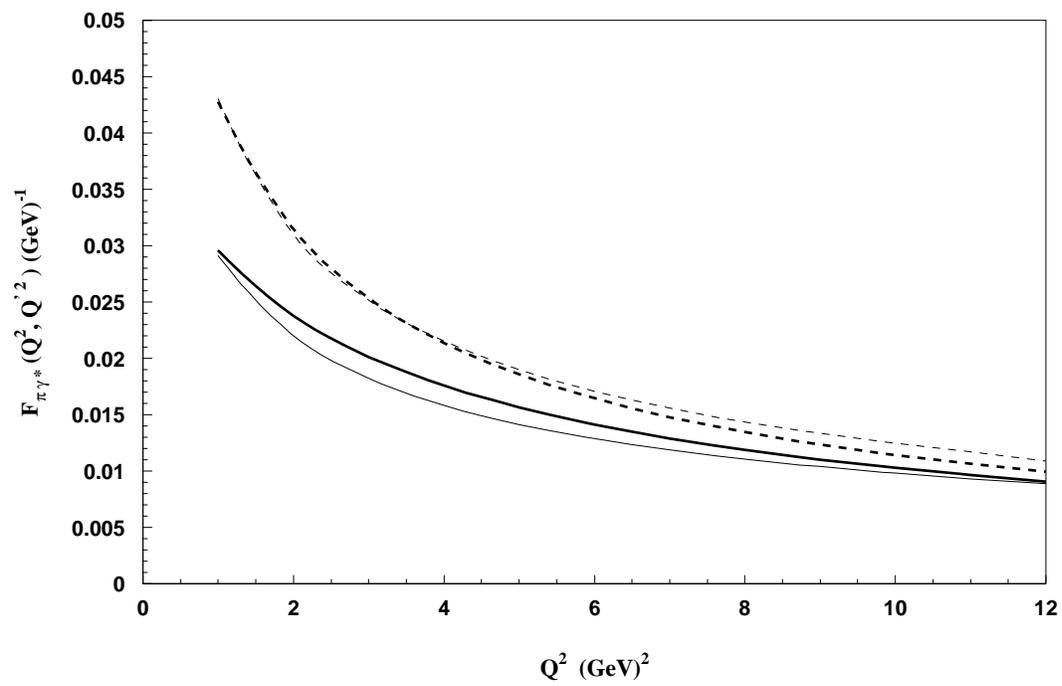
\item{Fig.~3} The $\pi$-$\gamma^\ast$ form factor at
$Q'^2=2$ GeV$^2$. The explanation of the curves is similar to Fig.~2.
\end{description}
\end{document}